\documentstyle[preprint,aps]{revtex}

\begin{document}
\title{A quantum critical point in the transverse field of Mn$_{12}$ system}
\author{Yongjin Jiang$^{1,2}$,Haijin$^{1,2}$ Li and Ruibao Tao$^{1}$}
\address{1. Department of Physics, Fudan University, Shanghai 200433, China 
\\
2. State Key Laboratory of Applied Surface Laboratory, Fudan University}
\date{\today }
\maketitle

\begin{abstract}
Using exact diagonalization method,we studied the ground state of the
anisotropic molecular magnets and find a critical point in the transverse
field,which may divide the quantum tunneling region into two different
parts. Possible ways to observe and take advantage of this point by varying
the transverse field are suggested.
\end{abstract}


Recently the macroscopic quantum phenomena in large spin anisotropic magnets
is attracting much attention among both theorists and
experimentalists[1-2].Many efforts has been done to measure the step-wise
magnetization loop,classify different temperature regimes, and find
different theoretical explanation for each. Quantum resonant tunneling is
the underlying mechanism that leads to abrupt change in the magnetsation.
However,it's not easy to take into account various factors that may
influence the result quantitatively,which sometimes results in sheerly
different pictures[3-6]. Fortunately,the model Hamiltonian has been
established and widely used in theoretical analysis. Many work has been
starting from the single large spin model, treating other parts as mean
fields,with some static or dynamic distribution function. There are three
different temperature regimes that can be described as different
phases:thermal-activated regime,thermal-assisted regime,and purely quantum
tunneling regime. Apparent phase transition(PT) has been studied by
analytical and numerical means and both first and second order phase
transition are found[7]. In this paper,through a properly defined order
parameter, we find a kind of quantum phase transition(QPT) for Mn$_{12}$
which is induced by continuous change of transverse field in the third
regime. Its experimental consequences is suggested.

Firstly,the model Hamiltonian of Mn$_{12}$ is

\begin{equation}
H=-D(S^z)^2-B(S^z)^4-C((S^{+})^4+(S^{-})^4)-g\mu _Bh_xS^x-g\mu _Bh_zS^z
\label{0}
\end{equation}
\index{1}

where the first and second terms are diagonal part of the crystal field,the
third term is the off-diagonal part. The last two terms represent external
field. In writing down it, we neglect all the other internal degrees of
freedom of the cluster on one lattice site,i.e.,the high energy spectrum
which can be deviating the description of fixed spin magnitude,for the
antiferromagnetic coupling between internal spins (4 spins 3/2 and 8 spins 2
for Mn$_{12})$ is rather large. This is called the ``mesoscopic '' approach.
The parameters

are determined by various measurements. In the following we choose $D\approx
0.60K,B/D$ $\approx 0.002,$ $C\approx 3\cdot 10^{-5}K,$which are obtained in
EPR experiments. Furthermore we set $h_{z}=0$ to study pure transverse field
effect.

In the pure quantum relaxation regime,without the longitudinal
field,tunneling occurs between degenerate ground states $-S$ and $+S.$ In
the Landau-Zener theory the transition probability is given by:

\begin{equation}
P_{-S,S}=1-\exp (-\Gamma _{-S,S})  \label{1}
\end{equation}

with 
\begin{equation}
\Gamma _{-S,S}=\pi \Delta _{(-S,S)}^2/[4\hbar Sg\mu _BdH/dt]  \label{2}
\end{equation}

which is determined by tunneling splitting as well as sweeping rate of
applying field. In this paper,however, we will study a related problem---the
characteristic of the ground state--- and understand the physics in a
different way. We consider the states with largest $\left| \langle
S_z\rangle \right| $,denoted as $\left| g\right\rangle ,$which is also
energetically lowest in most cases. Generally $\left| g\right\rangle $ can
be expanded as 
\begin{equation}
\ \ \ \left| g\right\rangle =\sum\limits_{m=-S}^Sc_m\left| m\right\rangle
\label{3}
\end{equation}

Now we define order parameter as the difference of probability of the $%
\left| \pm S\right\rangle $ components in $\left| g\right\rangle :$

\begin{equation}
\delta =\left| c_s\right| -\left| c_{-s}\right|  \label{4}
\end{equation}

\strut For situations where quantum tunneling due to ground state dynamics
dominates in sufficiently low temperature, $\delta $ measures the ability of
ground state $\left| g\right\rangle $ to transfer spins from one side in
magnetic spectrum to other side. Some cases corresponding to $%
(1)B=0,C=0.(2)B=0.002,C=0.(3)B=0,C=10^{-5}$ ,are studied and demonstrated in
Figure 1.The $\delta $ line changes abruptly near some point of $h_x$%
,showing a quantum critical point $h_c$ in the classical limit $%
S\longrightarrow \infty $.Across this narrow crossover region,the difference
of $\left| \pm S\right\rangle $ components in $\left| g\right\rangle $(i.e.,$%
\delta $)$,$changes from almost 1 to 0.When $\left| h_x\right| >h_c,$
numerical results show equal components of the ``bare'' states $S=\pm 10,$in
our case. The effect of $B$ and $C$ term on the $\delta $ line,especially on
the crossover position are easy to understand.$B$ term tends to strengthen
the anisotropy,leading to higher barrier between the positive and negative
half of the magnetic spectrum. Thus a larger value of $\left| h_x\right| $
is needed to cause tunneling combination. However, the $C$ term is
transverse and will help tunneling,and because of its four-ordered operator
form,can influence the results greatly by a little change of its value($%
10^{-5}$ relative to $D$,and $10^{-2}$ to $B$). The combined effect is shown
in Figure 2.The $C$ term quickly decreases the difference of the
two``phases''. In some systems, $C\approx 3\cdot 10^{-5}$ so maybe we need
some trick to really observe the critical point.

Also we calculated the distribution on bare states of $\left| g\right\rangle 
$.As is evident from Figure 3.The $\left| \pm S\right\rangle $ components
are sizable near the crossover region in the $\delta =0$ phase. The
experimental consequence is,in a well prepared sample,if we drive all sample
sites to $\left| S=10\right\rangle $ state,a considerable portion of lattice
site can tunnel to the $\left| S=-10\right\rangle $ state through the ground
state channel, if the transverse field is properly controlled. Or,by
increasing the transverse field across the crossover region,we can ``pump''
many sites from one side of the magnetic spectrum to another. Such scenario
is related to relaxation experiment\cite{relaxation}.

To conclude,by means of exact diagonalization,we find a crossover in the
Hamiltonian of molecular magnets,through which the difference of amplitudes
of $\left| \pm S=10\right\rangle $ states in ground state are changed from
almost 1 to zero. This property might have some experimental consequences in
the purely quantum regime of the magnetization relaxation which is caused by
tunneling through the ground state channel.

\bigskip

{\Large Acknowledgment}

This work is supported by National Natural Science Foundation of China,
Shanghai Research Center of Applied Physics and Institute of Physics of
Chinese Academy of Sciences.

\bigskip \newpage

{\LARGE Caption:}

{\normalsize Figure 1. }The amplitude difference $\delta $ of the $\left|
\pm S\right\rangle $ components in the ground state $\left| g\right\rangle $
for three groups of $B,C$ values.The $\delta $ line changes abruptly near
some $h_x$ points. The effect of $B$ is to broaden the cental ``phase''of
the diagram, while the $C$ value effects the opposite way.

\bigskip

{\normalsize Figure 2. }For larger $C$ values the central phase narrows
quickly and will disappear at large enough $C$ value.This made observation
of the critical point of transverse field more subtle and difficult.Here we
plotted two lines corresponding to $C=10^{-5}$ and $C=2\cdot 10^{-5}.$

\bigskip

{\normalsize Figure 3.}The amplitudes of the ground state projected to bare
states:$\left| -S\right\rangle ,...,\left| S\right\rangle $,for cases $(a)$ $%
h_{x}=1,(b)$ $h_{x}=2.3$ ,$(c)$ $h_{x}=2.7,(d)$ $h_{x}=3.3.$ Both $B$ and $C$
are choosen to be zero.The distribution tends to become symmetric as $h_{x}$
transcends some value.


\bigskip

\bigskip

\begin{references}
\bibitem{Barbara review1} B.Babara, L.Thomas, F.Lionti, I.Chiorescu,
A.Sulpice, J.Magn.Magn.Mat {\bf 200}, 167(1999);

\bibitem{Barbara review2} B.Babara, I.Chiorescu, R.Giraud, A.G.M.Jansen,
A.Caneschi, cond-mat/0005268.

\bibitem{phonon} G.Bellessa,N.Vernier,B.Barbara, and
D.Gatteschi,Phys.Rev.Lett {\bf 83},416(1999).

\bibitem{hyperfine} D.A.Garanin,E.M.Chudnovsky,and R.Schilling,Phys.Rev.B%
{\bf \ 61,}12204(2000);

\bibitem{Dislocation} E.M.Chudnovsky and D.A.Garanin,Phys.Rev.Lett {\bf 87}%
,187203-1(2001).

\bibitem{spin Shuffling} Jie Liu,Biao Wu,Li-Bin Fu,R.B.Diener and Qian
Niu,cond-mat/0105497(2001).

\bibitem{numerical PT} See,D.A.Garanin,E.M.Chudnovshy,Phys.Rev.B {\bf 63,}%
024418-1(2000).{\bf \ }

\bibitem{relaxation} W.Wernsdorfer, et al., Phys.Rev.Lett. {\bf 82}, 3903
(1999); N.V.Prokof'ev and P.C.E.Stamp, Phys.Rev.Lett. {\bf 80}, 5794 (1998).
\end{references}
\end{document}